# Quantum Hall effects in two-dimensional electron systems: A global approach

M. A. Hidalgo

Departamento de Física y Matemáticas, Facultad de Ciencias, Universidad de Alcalá, Alcalá de Henares (Madrid), Spain

Electronic mail: miguel.hidalgo@uah.es

**Abstract**
Up to almost the last two decades all the experimental results concerning the quantum Hall effect (QHE), i.e., the observation of plateaux at integer (IQHE) or fractional (FQHE) values of the constant $h/e^2$, were related to quantum-wells in semiconductor heterostructures (QWs). However, more recently, a renewed interest in revisiting these phenomena has arisen thanks to the observation of entirely similar effects in graphene and topological insulators. In this paper we show an approach encompassing all these QHEs using the same theoretical frame, entailing both Hall effect plateaux and Shubnikov-de Haas (SdH) oscillations. Moreover, the model also enables the analysis of both phenomena as a function not only of the magnetic field but the gate voltage as well. More specifically, in light of the approach, the FQHE in any two-dimensional electron system (2DES) appears to be an effect of the breaking of the degeneration of every Landau level, $n$, as a result of the electrostatic interaction involved, and being characterized by the set of three integer numbers ($n$, $p$, $q$), where $p$ and $q$ have clear physical meanings too.

## 1. Introduction

The discover of the integer quantum Hall effect in 1980, [1], and the fractional quantum Hall effect two years later in 1982, [2], in a two-dimensional electron system (2DES) embedded in a semiconductor quantum-well (QW), were a milestone for both, their intrinsic theoretical interest and their potential technological implications. Later on the observation of similar phenomena in graphene relaunched the interest on the magnetotransport properties of 2DES. Actually, these are alike in any 2DES: minima, or eventually zeroes, in the longitudinal resistivity, i.e. Shubnikov-de Haas effect (SdH), and plateaux in the Hall resistivity at integer or fraction values of the fundamental Hall resistance, $R_H = h/e^2$. Despite the similarities amid the quantum Hall effects (QHE) there are some differences: while the IQHE in QW semiconductors shows plateaux at integer values $n$=1, 2, 3, 4… (at $2n$ values if there is no broken spin degeneration), [3], the plateaux in monolayer graphene (MLG) are observed at the integers ±2, ±6, ±10, ±14, ±18, ±22…, i.e., following the sequence ±2(2n+1), [4] and [5].

Additionally, although the integer phenomenon is observed in any 2DESs confined in QW, either MOSFET or semiconductor heterostructures, high motilities heterostructures (like GaAs-AlGaAs) are needed to be the fractional phenomenon obtained. As a matter of fact, the appearance of fractional plateaux was completely unexpected. Firstly, the 1/3 plateau was measured, [2], and afterwards a large series of other fractional plateaux: {4/5, 2/3, 3/5, 4/7, 5/9, 5/3, 8/5, 11/7, 10/7, 7/5, 4/3, 9/7}; {5/3, 8/5, 10/7, 7/5, 4/3, 9/7, 4/5, 3/4, 5/7, 2/3, 3/5, 4/7, 5/9, 6/11, 7/13, 6/13, 5/11, 4/9, 3/7, 2/5, and also 8/3, 19/7, 33/13, 32/13, 7/3, 16/7}; {2/3, 7/11, 3/5, 4/7, 5/9, 6/11, 5/11, 4/9, 3/7, 2/5, 1/3, 2/7, 3/11, 4/15, 3/13, 2/9}; {14/5, 19/7, 8/3, 13/5, 23/9, 22/9, 17/7, 12/5, 7/3, 16/7, 11/5}; {19/5, 16/5, 14/5, 8/5, 7/3, 11/5, 11/3, 18/5, 17/5, 10/3, 13/5, 12/5}, [6]. A remarkable fact in the experiments is that most of the plateaux has odd denominator, while appear many fewer fractions with even denominator: {15/4, 7/2, 13/4, 11/4, 5/2, 9/4}; {11/4, 5/2, 9/4}; {1/4}; {11/4, 21/8, 5/2, 19/8, 9/4}; {7/2, 5/2}; {7/2, 5/2, 3/2, 1/4, 3/4}, [6]. Moreover, this phenomenon is not unique of QW but also observed in graphene, [7].

Analysing and comparing the extensive plateaux series and their recurrence in the experiments on different 2DESs (QWs, graphene…), the fundamental character not only of the IQHE but also the FQHE phenomena is clear, leading inevitably to think in a common origin of the physics underneath.

From the theoretical point of view whereas the IQHE is considered to be a consequence of the single particle localization effect under quantum limit conditions, [8], the strong electron correlations in partially filled Landau level is believed to be after the FQHE. In fact Laughlin proposed the formation of a correlated incompressible electron liquid as the origin of any $1/q$ plateaux, $q$ odd integer, [9] and [10]. This approach is based in the search of *ad hoc* trial wavefunction involving electron-electron interaction and corresponding to states that do not break the continuous spatial symmetry and show energy gaps. However, after the observation of many other fractional plateaux, now globally gathered in the expression $p/(2sp±1)$, $s$ and $p$ integers, it was clear the necessity of addressing the issue from other point of view, and that was the reason for introducing a new concept, the composite fermions: quasiparticles formed by the electrons capture of an even number of flux quanta. Then, following this idea, the FQHE is seen as an integer quantum Hall effect of those quasiparticles, [11].

Therefore, both phenomena, IQHE and FQHE, are theoretically depicted in different scenarios (electron localization and transitions to a Fermi liquids, respectively), however, several crucial points remain unexplained, such as the fact that in all FQHE measurements the integer plateaux are also observed and, moreover, with much better resolution the higher the 2DES mobility is; the disproportion between the number of odd and even denominator fractions observed; the appearance of the FQHE in graphene, etc. All these facts compelled us to revise the theoretical explanations of both phenomena and to look for a different approach.

Nowadays, there are many recent, and historical, good reviews describing and analysing in detail all the experiments and models involving the phenomena we are interested in, ([12], references therein).

The structure of the paper is the following: in section 2 we highlight an experimental fact related to the maxima of the SdH oscillations that becomes important for the developing of the model proposed. Section 3 is devoted to the determination of the density of states of the 2DESs under QHE conditions. This magnitude is required to the calculation of both magnetotransport magnitudes: the longitudinal and transversal conductivities, detailed in Section 4. In Section 5 we show some examples of the results obtained with the model for the different QHEs. Eventually, Section 6 sums up the paper and includes the conclusions.

## 2. The maxima of the experimental SdH oscillations

It is well-known that the electron density of any 2DES, $n_0$, can be experimentally measured from Hall measurements at low magnetic field, from where the Fermi level is easily determined through the relation

$$E_F = \frac{2\pi\hbar^2 n_0}{m^*} \tag{1}$$

$m^*$ being the electron effective mass. On the other hand, from the experimental measurements of the SdH effect, we prove that the maxima of the corresponding oscillations are observed at the values of the magnetic field given by the expression

$$B_{SdH\,max} = \frac{m^* E_F}{e\hbar\beta} \tag{2}$$

where the factor $\beta$ depends on the QHE: $\beta = (n+1/2)$ for the IQHE in QWs semiconductors; $\beta = n$ for the IQHE in graphene; and $\beta = p/q$ for the FQHE in QWs semiconductors, as well as in graphene. In all these expressions $n$, $p$ and $q$ are integer, although $q$ odd (see below).

## 3. Density of states for the different 2DES under quantum Hall conditions:

Firstly we have to determine the energy levels for each electron of the 2DES under the application of an external magnetic field. (In all below we will also assume the effective mass approximation.)

### *3.1. IQHE in QWs semiconductors*:

The initial Hamiltonian corresponding to the first approach usually used for the study of this problem is given by the expression

$$H_1 = \frac{1}{2m^*}\left(\vec{p} + e\vec{A}\right)^2 \tag{3}$$

In all of our analysis we choose the symmetric gauge: $\vec{A} = B(-y,x)/2$, being $\vec{B} = (0,0,B)$ the applied magnetic field, assumed to be perpendicular to the plane defined by the electron system.

The possible energy states obtained from Eq. (3) correspond to the Landau levels

$$E_n = \left(n + \frac{1}{2}\right)\hbar\omega_c = (2n+1)E_0 \tag{4}$$

$n=0, 1, 2\ldots$, $\omega_c = eB/m^*$ the angular frequency, and $E_0 = \hbar\omega_c/2$. On the other hand, the wave functions are

$$\psi_n^m = \left[2^m \frac{(m+n)!}{n!}\right]^{-\frac{1}{2}} (x' - iy')^m \exp\left(-\frac{r'^2}{4}\right) L_n^m\left(\frac{r'^2}{2}\right) \tag{5}$$

with $x' = x/R$, $y' = y/R$, $r'^2 = x'^2 + y'^2$, $R = \sqrt{\hbar/eB}$, the magnetic length, and $L_n^m(r'^2/2)$ the Laguerre polynomials. These wave functions are orthonormal each other respect to both, the $n$ and $m$ indexes, the last one associated with the angular momentum of each electron. The energy levels given by Eq. (4) are all degenerated in $m$.

The waves functions, Eq. (5), can be spatially envisioned like cyclotron orbits, characterized by the Larmor radius, $R_O$, whose expected value can be determined from its square through the expression, [13],

$$<R_O^2> = qR^2 \tag{6}$$

$q$ being an odd number.

For the determination of the electron magnetotransport properties of the 2DES, we have to obtain the density of states from Eq. (4). Using the Poisson sum formula, we can get for it the expression, [14],

$$g(E) = g_0\left\{1 + 2\sum_{p=1}^{\infty} A_{\Gamma,p} A_{s,p} \cos\left[2\pi p\left(\frac{E}{\hbar\omega_c} - \frac{1}{2}\right)\right]\right\} \tag{7}$$

where the $A_{\Gamma,p}$ term is associated with the width of the Landau levels on account of the interaction of electrons with defects and impurities. For the sake of simplicity we assume gaussian width for each energy level, with a width, $\Gamma$, independent of the magnetic field, what entails that, [15],

$$A_{\Gamma,p} = \exp\left\{-\frac{2\pi^2 p^2 \Gamma^2}{\hbar^2 \omega_c^2}\right\} \tag{8}$$

The other additional term in the sum, $A_{S,p} = cos\left(\pi p \frac{g^*}{2} \frac{m^*}{m_0}\right)$, the Zeeman term, takes into account the effect of the magnetic field over the spin of electrons. $g^*$ is the generalized gyromagnetic factor, that we here suppose to be 2, and $m_0$ the free electron mass.

### 3.2. FQHE in QWs semiconductors:

In this section we present the deduction of the energy levels of electrons under fractional quantum Hall conditions. For it we take into consideration that any 2DES is embedded in a more complex structure, what necessarily means involving the electrostatic interaction with ionized impurities, i.e., we have to analyse the Hamiltonian

$$H_2 = \frac{1}{2m^*}\left(\vec{p} + e\vec{A}\right)^2 + U(\vec{r}) \tag{9}$$

Anyway, in all below, we suppose that this term is a perturbation with regard to the main term of the magnetic field.

At low magnetic fields the Larmor radius, Eq. (6), is large, and the effect of the electrostatic term over every electron can be assumed as spatially uniform, i.e., $<U(\vec{r})> \approx U_0$, and then the energy states given by the $E_n = (2n+1)E_0 + U_0$. However, at high magnetic fields the Larmor radius becoming smaller and smaller and the electrostatic potential turns into a non-uniform term, $U(\vec{r})$. Therefore, this means a change in the symmetry of the Hamiltonian $H_2$, Eq. (9), as the magnetic field augments: from the continuous spatial symmetry to a "discrete" one associated with the distribution of impurities, this characterized by a mean distance $d_i$, value ultimately determined by the density of impurities.

Before going ahead seeking the new electron energy states under fractional quantum Hall conditions, we have to capitalize on the concept of cyclotron trajectory, and in particular of its Larmor radius, (Eq. 6): This enables us to consistently establish the wavelength of the electron through the relation $\lambda_q = 2\pi\sqrt{<R_O^2>}$, and its corresponding wave number by

$$k_q = \frac{2\pi}{\lambda_q} = \frac{1}{\sqrt{<R_O^2>}} = \frac{1}{\sqrt{q}}\frac{1}{R} = \frac{1}{\sqrt{q}}\sqrt{\frac{eB}{\hbar}} \tag{10}$$

that will be crucial in our analysis.

We have seen above that at high magnetic fields $d_i \geq 2R_O$, and then the electrostatic interaction with the ionized impurities restricts the locations of electrons, characterized by correlations lengths $\eta d_i$, with $\eta$ fixed by the own distribution of impurities (see below). Hence, under this condition, the Hamiltonian $H_2$ is required to verify the relationship

$$H_2(r) = H_2(r + \alpha \eta d_i) \tag{11}$$

being $\alpha$ integer, and the cyclotron orbits to present the corresponding spatial correlation

$$<\psi_n^m(r + \alpha \eta d_i) / H_2 / \psi_n^m(r + [\alpha \pm 1]\eta d_i)> = \pm \gamma/2 \tag{12}$$

(In all below we despise higher order correlations and suppose that $\gamma$ is the same for any cyclotron orbit.) Although the set of wave functions $\{\psi_n^m\}$, (Eq. 5), does not verify the Bloch theorem expressed in Eq. (11), from them we can develop a new base making a linear combination of those wave functions as follows

$$\Phi_{d_i}(K,r) = \frac{1}{\sqrt{N}}\sum_{\alpha=1}^{N} exp(iK \cdot \alpha \eta d_i)\psi_n^m(r - \alpha \eta d_i) \tag{13}$$

$N$ is the normalization term corresponding to the number of cyclotron orbits in the 2DES, and $K$ the wave number associated with the new spatial symmetry. This equation represents an ensemble of orthonormal functions, such that $\Phi_{d_i}(K, r + \beta \eta d_i) = exp(iK \cdot \beta \eta d_i)\Phi_{d_i}(K,r)$, and $E = <\Phi_{d_i}(K,r) / H_2 / \Phi_{d_i}(K,r)>$.

And, therefore, the energy states are given by the equation $E = E_n \mp \gamma cos(K\eta d_i)$.

But the solutions of the Hamiltonian $H_2$, Eq. (9), at high magnetic fields have to involve that $K=k_q$, with $k_q$ given by Eq. (10), and then $E = E_n \mp \gamma cos(k_q \eta d_i)$. Even more we can approach this equation by this other one $E = E_n \mp \gamma \left[1 - \frac{1}{2}\left(k_q \eta d_i\right)^2\right]$. Comparing this with the result obtained at low magnetic field, it is easily to deduce that $\gamma = U_0$, and then

$$E = E_n \mp U_0 \pm \frac{U_0}{2}\left(k_q \eta d_i\right)^2 \qquad (14)$$

This equation is the energy of a quasi-free electron with effective mass $m^* = \hbar^2 / |U_0| d_i^2$. Using Eq. (4) and Eq. (10) in Eq. (14) we obtain

$$E = E_n \pm \frac{\eta^2}{q} E_0 \pm U_0 = \left(2n + 1 \pm \frac{\eta^2}{q}\right) E_0 \pm U_0 \qquad (15)$$

where $E_0 = \hbar \omega_c / 2$. The most probable expected correlations in the arrangement of the electron cyclotron orbits in the 2DES due to the electrostatic potential correspond to values of the correlation length parameter of $\eta=1$ and $\sqrt{3}$, [16], just like in a Wigner crystal. In general we will express this parameter as $\eta = \sqrt{p}$. Thus, the dependent part of the magnetic field of Eq. (15) can be rewritten as

$$\frac{E}{\hbar \omega_c} = \frac{1}{2}\left(2n + 1 \pm \frac{p}{q}\right) \qquad (16)$$

Therefore, the fractional energy states are defined by the three factors ($n$, $p$, $q$), and in Table I we detail the states for the Landau levels $n=0$, 1 and 2; $p=1$ and 3, i.e., correlations lengths $\eta=1$ and $\eta=\sqrt{3}$, respectively; and $q$ values between 3 and 13.

Special interest deserves the states related to the odd number $q=1$, which corresponds to a Larmor radius equal to the magnetic length, Eq. (6). Then, if the electron arrangement in the 2DES has a correlation length parameter like $\eta = 1/\sqrt{j}$, [16], Eq. (16) provides us the energy states

$$\frac{E}{\hbar \omega_c} = \frac{1}{2}\left(2n + 1 \pm \frac{1}{j}\right) \qquad (17)$$

In Table II we depict the energy states for Landau levels $n=0$, 1 and 2, and $j=2$ and 4, i.e., correlations lengths with $\eta = 1/\sqrt{2}$ and $\eta=1/2$.

The last point related to Eq. (16) concern with the energy states with correlation index $p=0$, what means lacking of the spatial correlation of the electron.

Once we have deduced the energy states we can calculate the density of states of the 2DES for every fraction series of the FQHE. Following the same steps as used to determine Eq. (7), we can obtain for any $q$ index, [14],

$$g_q(E) = g_0 \left\{1 + 2 \sum_{p=1}^{\infty} A_{\Gamma,p} \cos\left[\left(\frac{2\pi p q E}{\hbar \omega_0} - \frac{g^*}{4} \frac{m^*}{m_0}\right)\right]\right\} \qquad (18)$$

and where the meaning of each factor is the same as in Eq. (7).

### 3.3. IQHE in monolayer graphene:

The observation of the IQHE and FQHE in monolayer graphene (MLG) awakened again the interest in both phenomena. And we have found that the tools developed in the previous subsections for these phenomena in QWs semiconductors can be easily extended to encompass both effects in graphene. The crucial feature for this purpose is its semimetal character and the fact of having a zero-band gap with the Fermi level located at the intersection between the valence and the conduction bands, [17, therein]. Therefore, under the application of a magnetic field the electron states will be given by the superposition of the dynamical spaces of both bands, what entails two degrees of freedom, and consequently the energy spectrum correspond to that of an isotropic bi-dimensional harmonic oscillator, i.e.,

$$E_n = (i + j) \hbar \omega_c = n \hbar \omega_c \qquad (19)$$

with $i$ and $j$ natural numbers related to each dynamical space. In the same way as we obtained the density of states for QWs semiconductors, Eq. (7), we can now determine the corresponding one for gaphene, [18],

$$g(E) = g_0 \left\{ 1 + 2 \sum_{p=1}^{\infty} A_{\Gamma,p} A_{s,p} \cos\left[2\pi p \left(\frac{E}{\hbar\omega_c}\right)\right] \right\} \tag{20}$$

the same for both bands. And, then, the total density of states will be

$$g^{total}(E) = g^{CB}(E) + g^{VB}(E) \tag{21}$$

### 3.4. FQHE in monolayer graphene:

It is predictable that the origin of the FQHE in graphene is the same as in QWs semiconductors. But now we have to take into account the two-dimensional dynamical space, what implies that the wave number has to be given by

$$k_q = \sqrt{\left(k^{CB}\right)^2 + \left(k^{VB}\right)^2} = \sqrt{2} k_q^{VB} = \sqrt{2} \frac{2\pi}{\lambda_q} = \sqrt{\frac{2eB}{q\hbar}} \tag{22}$$

assuming that each band component of the wave number is the same.

Therefore, following the same steps as in Subsection 3.2, although now considering Eq. (19), we obtain the expression for the electron energy states in graphene under fractional quantum Hall conditions, which assuming a correlation length $\eta=1$ is given by

$$\frac{E}{\hbar\omega_c} = \left(n \pm \frac{1}{q}\right) \tag{23}$$

In Table IV we detail the energy states for Landau levels $n=0$, 1 and 2, and $q=3$, and a correlation length $\eta=1$. In this case the calculated density of states is similar to that obtain for the FQHE in QWs semiconductors, Eq. (18), [18].

### 4. The model for both magnetoconductivities:

Theoretically the magnetotransport properties of any 2DES are determined by the magnetocoductivities that at high magnetic fields have the expressions, [14], for the longitudinal one, the SdH effect:

$$\sigma_{xx} = \frac{1}{\omega_c \tau} \frac{en_0}{B} \frac{g(E_F)}{g_0} \tag{24}$$

where $g(E_F)$ is the density of states at the Fermi level, $E_F$, obtained from Eq. (1), $g_0$ the 2DES density of states at zero magnetic field and $\tau$ the relaxation time of the electrons. And, on the other hand, for the Hall magnetoconductivity we get

$$\sigma_{xy} = -\frac{en}{B} \tag{25}$$

where $n$ is the electron density in the 2DES, as obtained from the corresponding density of states.

From these Eq. (24) and Eq. (25), the symmetric magnetoresistivity tensor, $[\rho]=[\sigma]^{-1}$, can be calculated through the relations

$$\rho_{xx} = \rho_{yy} = \frac{\sigma_{xx}}{\sigma_{xx}^2 + \sigma_{xy}^2} \tag{26}$$

$$\rho_{xy} = -\rho_{yx} = -\frac{\sigma_{xy}}{\sigma_{xx}^2 + \sigma_{xy}^2} \tag{27}$$

### 5. QHE and SdH simulations:

This section is devoted to present the results of the model for every QHE analysed above. We indistinctly simulate the behaviour of both, magnetoconductivities or magnetorresistivities, and as a function of the gate voltage or the magnetic field.

### 5.1. IQHE in QWs semiconductors:

Analysing the experimental data of the SdH effect we can directly prove that there is a direct relation between each maximum of its oscillations, Eq. (2) with $\beta = (n+1/2)$, and the corresponding plateau of index $n$ in the IQHE measurements. And the maxima of those oscillations, in turn, are given by Eq. (4).

As it is evident from Eq. (25), we need to obtain the electron density in the 2DES to determine the Hall magnetocoductivity. Thus, using Eq. (7) we have, [14]:

$$n = n_0 + \frac{2eB}{h} \sum_{p=1}^{\infty} \frac{1}{\pi p} A_{S,p} A_{\Gamma,p} A_{T,p} sen\left[2\pi p \left(\frac{E_F}{\hbar \omega_c} - \frac{1}{2}\right)\right] = n_0 + \delta n \qquad (28)$$

$n_0$ is the density of electrons at zero magnetic field (or zero gate voltage), and $\delta n$ the fluctuation in the own electron density as a consequence of the quantized density of states, and the fixed by the whole system Femi level. The effect of the temperature that determines the occupation of every state is included in the term $A_{T,p} = z/senh(z)$, where $z = 2\pi^2 pkT/\hbar \omega_c$, where $k$ is the Boltzmann constant and $T$ the corresponding temperature.

Hence, once we have the density of states, Eq. (7), and the density of electrons in the 2DES, Eq. (28), we are able to simulate both magnetoconductivities, Eq. (24) and Eq. (25).

Several results of the model for the IQHE in QWs semiconductors, and its comparison with experimental measurements, are shown in detail in references [14], devoted to the analysis of the IQHE.

### 5.2. FQHE in QWs semiconductors:

In this case each experimental SdH maximum, Eq. (2) with $\beta = p/q$, is related to the plateau of index $p/q$ in the FQHE measurements. And the maxima of the SdH oscillations, in turn, are given by Eq. (16) and/or (17). As in the previous section, from the density of states, Eq. (18), we can obtain the electron density:

$$n = n_0 + \frac{2eB}{hq} \sum_{p=1}^{\infty} \frac{1}{\pi p} A_{\Gamma,p} A_{T,p} sen\left[\frac{2\pi q p E_F}{\hbar \omega_c} - \frac{g^*}{4}\frac{m^*}{m_0}\right] \qquad (29)$$

from where together with Eq. (18) allow us to simulate both magnetoconductivities under fractional quantum Hall conditions.

In Figure 1 we show an example of the simulations obtained, in this case for the odd denominator states $q=3$ and $q=5$, and the corresponding plateaux.

### 5.3. IQHE in monolayer graphene:

From the experimental data of the SdH effect we can see that there is a direct relation between each maximum of its oscillations, Eq. (2) with $\beta = n$, and the corresponding plateau of index $2(2n+1)$.

Following the previous sections, using now Eq. (20) and Eq. (21) we calculate the electron density in graphene

$$n^{total} = 2\left(n_0 + \frac{2eB}{h} \sum_{p=1}^{\infty} \frac{1}{\pi p} A_{S,p} A_{\Gamma,p} A_{T,p} sen\left[2\pi p \left(\frac{E_F}{\hbar \omega_c}\right)\right]\right) \qquad (30)$$

with $n_0$ being the density of electrons at zero magnetic field. Then, using Eq. (21) and Eq. (30) we obtain both magnetoconductivities. In Figures 2 and 3 we show the results for the simulation with the parameters detail in the own figure captions, some of them taken from the reference [19]. As it is seen the model reproduces the measurements showing plateaux at the expected values ±2, ±6, ±10, ±14, ±18, ±22…

### 5.4. FQHE in monolayer graphene:

Following a similar procedure as described in Subsection 5.2, we can simulate the phenomenon of the FQHE in graphene. In Figure 4 we show the results of the model obtained for it as a function of the gate voltage. Because we have found a direct connection between the energy states in fractional quantum Hall conditions, (given by Eq. (16), Eq. (17) and Eq. (23)), and the fractional plateaux, and the fact that in most of the experiments the observed plateaux correspond to the set related to $q=3$, we can conclude that the most probable correlation length in grapheme, under fractional quantum Hall conditions, is $\eta=1$.

## 6. Summary and Conclusions

In this paper we have described a global approach for the QHEs observed in QWs semiconductors and graphene. In fact we have shown that they can be understood under the same theoretical frame.

In particular, the phenomenon of the FQHE arises due to the breaking of the symmetry of the Hamiltonian, Eq. (11), as the magnetic field change: at low magnetic fields the distribution of electrons in the 2DES has a spatially continuous symmetry, while at high magnetic fields, a discrete symmetry emerges associated with the ionized impurities distribution. Additionally, in Table I and II we have detailed the fractional energy states obtained for different discrete distributions of electrons in the 2DES. From them it is possible to deduce that, for example, the set of fractions {1/3, 2/5, 4/9, 5/11, 6/13}, {2/3, 3/5, 4/7, 5/9, 6/11, 7/13})... , (the columns in Table I), are all a consequence of the breaking of the degeneration of the first Landau level, $n$=0. As a matter of fact, we have found a direct relation between the energy states and the corresponding plateau. In Tables I-IV, we have highlighted in red the fractional plateaux already observed in the experiments, and in black those expected to be observed. More recently the plateaux series {1/9, 1/7, 2/11, 2/13, 2/15, 2/17, 3/17, 3/19} have also been observed, and in the light of the model we can associated them with the set of the three numbers ($n,p,q$): (0,7,9), (0,5,7), (0,7,11), (0,9,13), (0,11,15), (0,13,17), (0,11,17), (0,13,19), respectively. Furthermore, from our approach to the FQHE we can substantiate why the observed imbalance in the experiments amid odd and even plateaux.

As it is seen in Table III, and mention in Subsection 3.2, in light of the model the even denominator fractions, $(2n+1)/2$, have their origin in the states representing lack of the correlations among electrons, i.e., a correlation index $p$=0. But, one important question that has always intriguing researchers is the case of the plateaux associated with the state ½, *a priori* expecting to be one of the most significant, but missed in the experiments. In the scenario described by the model, this state and its corresponding plateau cannot be observed because the magnetic field necessary to reach it entails so small Larmor radius that the distribution of electrons is entirely conditioned by the ionized impurities distribution.

In addition, in this paper we have shown that the QHEs observed in MLG are included in the approach, simply imposing considering the quantization of both bands, the VB and CB of the semimetal graphene. In particular in the case of the FQHE we show that it has the same origin as in QWs semiconductors.

We hope that the approach described in this paper can be considered as a useful tool in the understanding of the conduction mechanism and magnetotransport properties in any 2DES.

In a forthcoming paper we will extend the general approach described to analyse the magnetotransport properties of multilayers graphene and the topological insulators. In particular the case of bilayer graphene (two layers of atom carbons) has already outlined in reference [20].

## References

[1] Klitzing K. v, Dorda G and Pepper M 1980 *Phys. Rev. Lett.* **45** 494

[2] Tsui D C, Störmer H L and Gossard A C 1982 *Phys. Rev. Lett.* **48** 1559

[3] Prange R E and Girvin S M Editors 1990 The Quantum Hall effect, Springer-Verlag

[4] Novoselov K S, Geim, A K, Mozorov S V, Jiang D, Katsnelson M I, Grigorieva I V, Dubonos S V and Firsov A A, 1005 *Nature* **438** 197

[5] Zhang Y, Tan Y-W, Stormer H L and Kim P 2005 *Nature* **438** 201

[6] Clark R G, Nicholas R J, Usher A, Foxon C T and Harris J J 1986 *Surface Science* **170** 141; Willet R L *et al*. 1987 *Phys. Rev. Lett.* **59** 1776; Willet R L 1988 *Phys. Rev.* **B59** 7881; Du R R, Störmer H L, Tsui D C, Pfeiffer L N and West K W 1988 *Phys. Rev.* B70 2944; Pan W, Stormer H L, Tsui D C, Pfeiffer L N, Baldwin K W and West K W, 2003 *Arxiv*: 0303428v1; Choi H C, Kang W, Das Sarma S, Pfeiffer L M, West K W 2007 *Arxiv*: 0707.0236v2; Shabani J and Shayegan M 2010 *Arxiv*: 1004.09/9v1

[7] Dean C R, Young A F, Cadden-Zimansky P, Watanabe K, Taniguchi T, Kim P, Hone J and Shepard K L 2011 *Nature Physics* **7** 693; Bolotin K, Ghahan F, Schulman M D, Stormer H L and Kim P 2009 *Nature* **462**, doi:10.1038/nature085582; Du X, Skachko I, Duerr F, Luican A and Andrei E Y 2009 *Nature* **462**, doi:10.1038/nature08522; Gharari F, Zhao Y, Cadden-Zimansky P, Bolotin K and Kim P 2011 *Phys. Rev. Lett.* **106,** 046801

[8] Laughlin R B 1981 *Phys. Rev.* **B 23** 5632

[9] Laughlin R B 1983 *Phys. Rev. Lett.* **50** 1395

[10] Chakraborty T and Pietiläinen P 1988 The fractional Quantum Hall effect, Ed. Springer-Verlag

[11] Jain J K 1992 *Advances in Physics* **41** 105

[12] Ezawa Z F 2013 Quantum Hall Effects: Recent Theoretical and Experimental Developments, Ed. World Scientific

[13] Cohen-Tannoudji C 1977 *Quantum Mechanics*, Ed. John Wiley & Sons

[14] Hidalgo M A 1995 PhD Thesis; Hidalgo M A 1988 *Microelectronic Engineering* **43-44** 453; Hidalgo M A and Cangas R 2007, *Arxiv*: 0707.4371

[15] Ando T, Fowler A B, F. Stern F 1982 *Reviews of Modern Physics* **54** 437

[16] Hidalgo M A 2013 *Arxiv*: 1310.1787

[17] Castro Neto A H, Guinea F, Peres N M R, Novoselov K S and Geim, A K 2009 *Rev. Mod. Phys.* **81** 109; Goerbig M O 2011 *Rev. Mod. Phys.* **83** 1193; Das Sarma S, Adam S, Hwang E H and Rossi E 2011 *Rev. Mod. Phys.* **83** 407

[18] Hidalgo M A 2014 *Arxiv*: 1404.5537. doi: 10.13140/2.1.2006.5923; Hidalgo M A 2015 *Arxiv*: 1507.05023. doi: 10.13140/RG 213737.9684

[19] Tiras E, Ardali S, Tiras Tm Arslan E, Cakmakyapan S, Kazar O, Hassan J, Janzén E, Ozbay E 2013 *J. Applied Phys.* 113, 043708

[20] Hidalgo M A, Cangas R 2016 *Arxiv*: 1602.02631

**Table I: Fractional energy states for correlation indexes *p*=1 and *p*=3; i.e., correlations lengths $\eta$=1 and $\eta$=3:** Energy states for Landau levels *n*=0, 1 and 2, *q* values between 3 and 13, and *p*=1 and 3. The corresponding fractional plateaux already observed in the experiments are highlighted in red.

| $\frac{E}{\hbar\omega_0}$ | $\frac{1}{2}\left(2n+1-\frac{1}{q}\right)$ | | | $\frac{1}{2}\left(2n+1+\frac{1}{q}\right)$ | | | $\frac{1}{2}\left(2n+1-\frac{3}{q}\right)$ | | | $\frac{1}{2}\left(2n+1+\frac{3}{q}\right)$ | | |
|---|---|---|---|---|---|---|---|---|---|---|---|---|
| | *n*=0 | *n*=1 | *n*=2 | *n*=0 | *n*=1 | *n*=2 | *n*=0 | *n*=1 | *n*=2 | *n*=0 | *n*=1 | *n*=2 |
| *q*=3 | $\frac{1}{3}$ | $\frac{4}{3}$ | $\frac{7}{3}$ | $\frac{2}{3}$ | $\frac{5}{3}$ | $\frac{8}{3}$ | - | - | - | - | - | - |
| *q*=5 | $\frac{2}{5}$ | $\frac{7}{5}$ | $\frac{12}{5}$ | $\frac{3}{5}$ | $\frac{8}{5}$ | $\frac{13}{5}$ | $\frac{1}{5}$ | $\frac{6}{5}$ | $\frac{11}{5}$ | $\frac{4}{5}$ | $\frac{9}{5}$ | $\frac{14}{5}$ |
| *q*=7 | $\frac{3}{7}$ | $\frac{10}{7}$ | $\frac{17}{7}$ | $\frac{4}{7}$ | $\frac{11}{7}$ | $\frac{18}{7}$ | $\frac{2}{7}$ | $\frac{9}{7}$ | $\frac{16}{7}$ | $\frac{5}{7}$ | $\frac{12}{7}$ | $\frac{19}{7}$ |
| *q*=9 | $\frac{4}{9}$ | $\frac{13}{9}$ | $\frac{22}{9}$ | $\frac{5}{9}$ | $\frac{14}{9}$ | $\frac{23}{9}$ | - | $\frac{12}{9}$ | $\frac{21}{9}$ | - | $\frac{15}{9}$ | $\frac{24}{9}$ |
| *q*=11 | $\frac{5}{11}$ | $\frac{16}{11}$ | $\frac{27}{11}$ | $\frac{6}{11}$ | $\frac{17}{11}$ | $\frac{28}{11}$ | $\frac{4}{11}$ | $\frac{15}{11}$ | $\frac{29}{11}$ | $\frac{7}{11}$ | $\frac{18}{11}$ | $\frac{26}{11}$ |
| *q*=13 | $\frac{6}{13}$ | $\frac{19}{13}$ | $\frac{33}{13}$ | $\frac{7}{13}$ | $\frac{20}{13}$ | $\frac{32}{13}$ | $\frac{5}{13}$ | $\frac{18}{13}$ | $\frac{34}{13}$ | $\frac{8}{13}$ | $\frac{21}{13}$ | $\frac{31}{13}$ |

**Table II: Fractional energy states for correlation indexes *j*=2 and 4; i.e. correlations lengths with $\eta = 1/\sqrt{2}$ and $\eta$=1/2:** Energy states for Landau levels *n*=0, 1 and 2, and *j*=2 and 4. As in Table I, the corresponding fractional plateaux already observed in the experiments are highlighted in red.

| $\frac{E}{\hbar\omega_0}$ | $\frac{1}{2}\left(2n+1-\frac{1}{j}\right)$ | | | | $\frac{1}{2}\left(2n+1+\frac{1}{j}\right)$ | | | |
|---|---|---|---|---|---|---|---|---|
| | *n*=0 | *n*=1 | *n*=2 | *n*=3 | *n*=0 | *n*=1 | *n*=2 | *n*=3 |
| *j*=2 | $\frac{1}{4}$ | $\frac{5}{4}$ | $\frac{9}{4}$ | $\frac{13}{4}$ | $\frac{3}{4}$ | $\frac{7}{4}$ | $\frac{11}{4}$ | $\frac{15}{4}$ |
| *j*=4 | $\frac{3}{8}$ | $\frac{11}{8}$ | $\frac{19}{8}$ | $\frac{27}{8}$ | $\frac{5}{8}$ | $\frac{13}{8}$ | $\frac{21}{8}$ | $\frac{29}{8}$ |

**Table III: Fractional energy states for correlation index *p*=0:** Energy states associated with lack of large range order correlation amid electron cyclotron orbits.

| $\dfrac{E}{\hbar\omega_0}$ | $\dfrac{1}{2}(2n+1)$ | | | |
|---|---|---|---|---|
| | n=0 | n=1 | n=2 | n=3 |
| - | $\dfrac{1}{2}$ | $\dfrac{3}{2}$ | $\dfrac{5}{2}$ | $\dfrac{7}{2}$ |

**Table IV: Fractional energy states for correlation length *η*=1:** All the corresponding fractional plateaux have already been observed in the experiments.

| $\dfrac{E}{\hbar\omega_0}$ | $\left(n-\dfrac{1}{q}\right)$ | | | | $\left(n+\dfrac{1}{q}\right)$ | | | | |
|---|---|---|---|---|---|---|---|---|---|
| | n=1 | n=2 | n=3 | n=4 | n=0 | n=1 | n=2 | n=3 | n=4 |
| q=3 | $\dfrac{2}{3}$ | $\dfrac{5}{3}$ | $\dfrac{8}{3}$ | $\dfrac{11}{3}$ | $\dfrac{1}{3}$ | $\dfrac{4}{3}$ | $\dfrac{7}{3}$ | $\dfrac{10}{3}$ | $\dfrac{13}{3}$ |

**Figure 1:** Simulation with the model of the Hall magnetoconductivity and both magnetoresistivities for the odd denominators states *q*=3 and *q*=5.

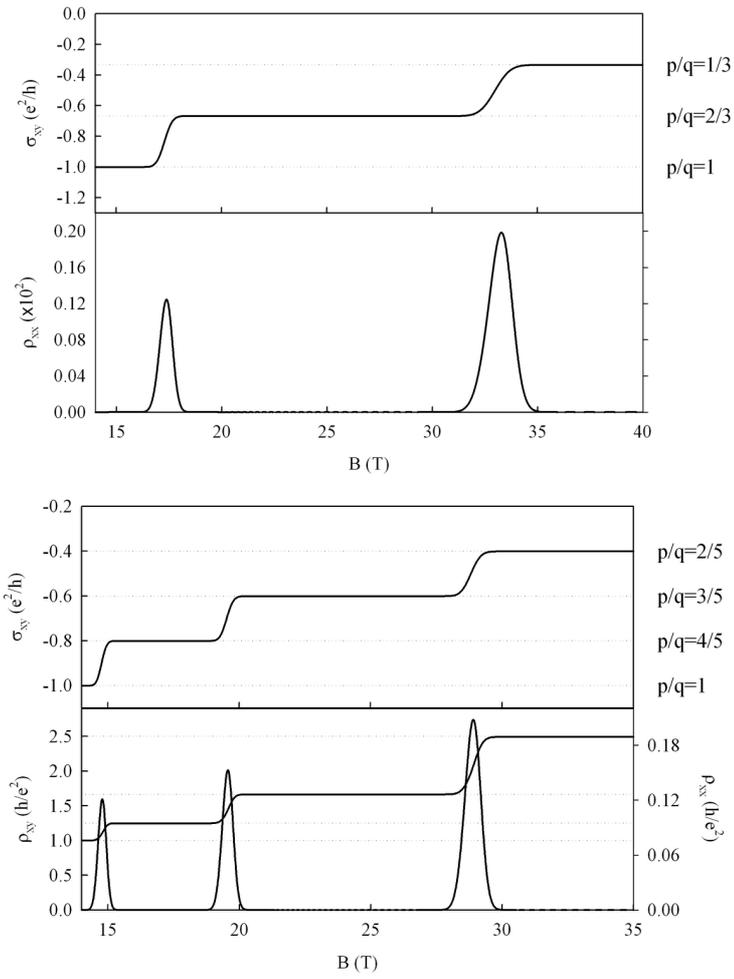

**Figure 2: Simulation in monolayer graphene of the Hall magnetoconductivity, (a), and Hall diagonal magnetoresistivities, (b), as a function of the gate voltage:** It has been obtained taking a temperature of 1.6 K, a magnetic field of $B=9$ T, a gyromagnetic factor of 2 and $n_0=0$. The effective mass used is $m^*=0.0124 m_0$, [19]. On the other hand, we considered a constant Gaussian width of $\Gamma=0.06$ eV and a relaxation time $\tau=1$ ps. In (a) the integer numbers over the reference lines label the different plateaux.

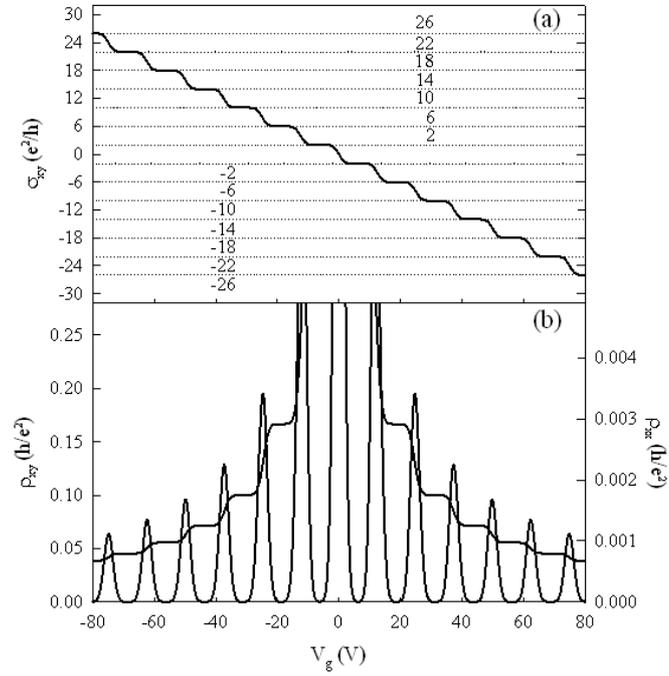

**Figure 3: Simulation in monolayer graphene of the Hall magnetoconductivity, (a), and Hall diagonal magnetoresistivities, (b), as a function of the magnetic field:** The electron density at zero magnetic field used in the simulation was $n_0=6\times10^{15}$ m$^{-2}$ and a Gaussian width of $\Gamma=0.025$ eV. The other parameters of the simulation were the same as in Figure 2.

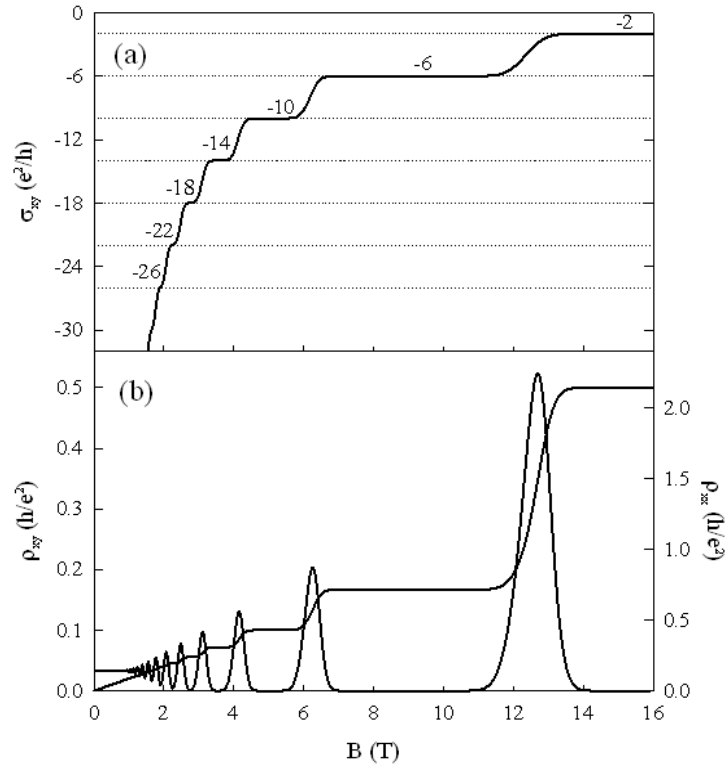

**Figure 4: Simulation in graphene of the fractional quantum Hall magnetoconductivity and both magnetoresistivities as a function of the gate voltage for the odd denominator q=3.**

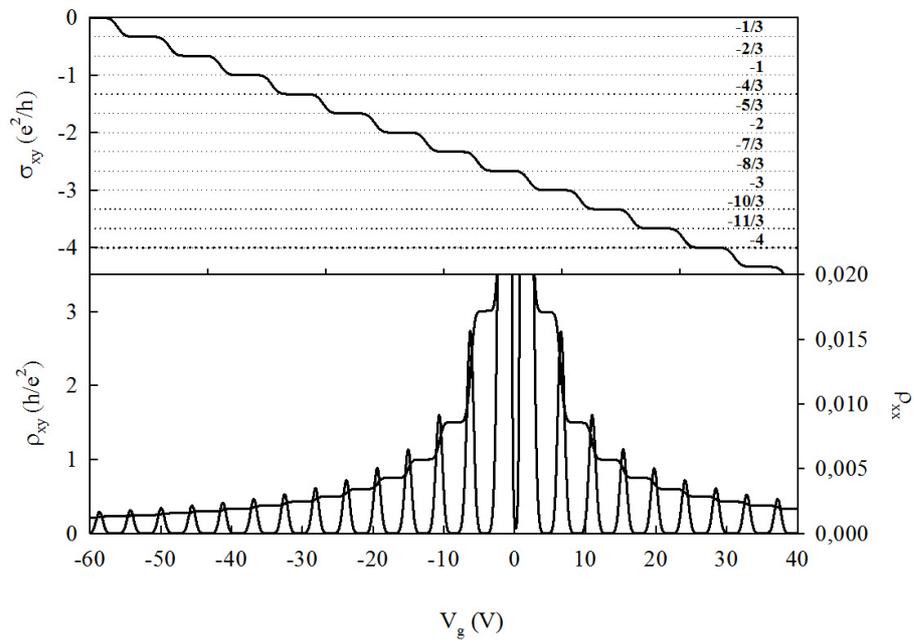